\documentclass[reprint,amsmath,amssymb,aps]{revtex4-1}
\usepackage{epsfig}
\usepackage{graphicx}
\usepackage{dcolumn}
\usepackage{comment}
\usepackage{color}
\usepackage{ulem}
\usepackage{longtable}
\usepackage{amsthm,amsmath}

\begin{document}

\title{Two-dipole and three-dipole interaction coefficients of GROUP XII elements}
\author{Neelam Shukla$^1$, Harpreet Kaur$^2$, Bindiya Arora$^2$\footnote{bindiya.phy@gndu.ac.in}, and Rajesh Srivastava$^1$\footnote{rajesh.srivastava@ph.iitr.ac.in}}
\affiliation{$^1$Department of Physics, Indian Institute of Technology Roorkee, 247667, Roorkee, India}
\affiliation{$^2$Department of Physics, Guru Nanak Dev University, Amritsar, Punjab-143005, India}

\begin{abstract}
The study of long-range interactions is increasingly becoming essential due to its various applications in cold atomic physics.  These interactions can be conveniently expressed in terms of dispersion coefficients. In the present work, theoretical calculations have been carried out to estimate the two-dipole ($C_6$)  and three-dipole ($C_9$) dispersion coefficients between the group XII atoms and their ions, \textit{viz.}, Zn, Cd, Hg, Zn$^+$, Cd$^+$, and Hg$^+$. To obtain these coefficients, the dynamic dipole polarizabilities and reduced matrix elements required are evaluated using the relativistic methods for both the considered atoms and ions. Further, using the calculated matrix elements, the oscillator strengths corresponding to leading transitions and static dipole polarizabilities are determined. These are compared with the available values in the literature through which the accuracy of the present calculations ascertains. However, the oscillator strengths for few transitions of Zn, Cd, Hg, Zn$^+$, Cd$^+$, and Hg$^+$, are not reported earlier, and we have provided their values in this work. Thereafter, the dispersion $C_6$ and $C_9$ coefficients are calculated for various combinations of the group XII elements, and most of these coefficients were not reported earlier in the literature. Furthermore, the combinations for which the other theoretical and experimental values were available for comparison showed good agreement. We, therefore, believe that our results reported for $C_6$ and $C_9$ coefficients for various combinations of group XII elements are reliable.

\end{abstract}

\maketitle

\section{Introduction}
 The research on laser-cooled ions and atoms is of great interest due to its involvement in several aspects of ultra-cold atomic physics. It has been suggested in several reports that the study of controlled ion-atom cold collisions can be utilized in future quantum information processing \cite{doerk2010,harter2014}. For this purpose, to gain the precise information about the long-range interatomic interactions among the cooled atoms and ions is very important because it plays key role in diverse investigations, \textit{viz.}, atom-ion hybrid traps, determination of stability of Bose-Einstein condensates \& scattering length, experiments on photoassociation, analysis of Feshbach resonances, and many other \cite{smith2005,zipkes2010,gacesa2016,roberts1998,amiot2000,leo2000,leanhardt2003,yuju2004}. The involvement of long-range interactions in low-temperature collisions \cite{zipkes2010,schmid2010,bodo2008,zhang2009,zhang20091,idziaszek2009,gao2010,cote2000,cote20001,makarov2003} make it essential to study and it also helps in determining the collisional frequency shift. Further, the investigation of such interactions provides fundamental knowledge about  charge transfer processes \cite{cote2000,cote20001,zhang2009,zhang20091,sayfutyarova2013}, ion-atom bound states \cite{cote2002}, formation of cold molecular ions, and spin exchange reactions \cite{schneider2010,staanum2010}, which helps us to understand them more clearly. 
 
 The long-range interaction potential is generally written in terms of the inverse of the interatomic distance $R$ as a power series (see Eq. (1)). The multiplier of the term that has $R^{-6}$ represents the two-body dipole interaction known as $C_6$ dispersion coefficient and contributes the most in the series, due to which it becomes imperative. Similarly, the term with $C_9$ dispersion coefficient acts as a leading contributor to the non-additive part of the potential when three-body dipole interaction is considered. Thus, making these two coefficients essential to explore.  Due to this reason, many theoretical and experimental studies have been reported considering various ion-atom combinations, and details of which are given in Ref. \cite{shukla2020}. In addition, different experiments have been performed to understand the nature of force in-between  the ultracold ions and atoms in hybrid traps \cite {smith2005,grier2009,zipkes2010cold, zipkes2010trapped,hall2011light,hall2012millikelvin,sullivan2012,ravi2012,lee2013,tacconi2011,mclaughlin2014}.
 
The elements of group XII (Zn, Cd, Hg) are very important because these are proposed as potential candidates for optical lattice clock \cite{wang2007,brickman2007,hachisu2008}. Further, elements from group II (alkaline-earth), group XII, and group XVIII (rare-gas) are considered very suitable for van der Waals molecules 
 \cite{qiao2012}.
In this regard, the dispersion coefficients of alkaline-earth and rare-gas dimers have been widely studied, and accurately determined \cite{porsev2002,shukla2020,mitroy2007,tang2010}. However, less attention has been paid towards the dispersion coefficients of group XII dimers, and only limited and sporadic data are available for immediate use. Therefore, this lack of data motivates us to calculate the two-dipole and three-dipole interactions for various combinations of group XII elements (Zn, Cd, Hg) systematically through the \textit{ab-initio} method. The $C_6$ coefficients are evaluated for atom-atom (homonuclear and heteronuclear pairs) and atom-ion dimers. In addition, the $C_9$ coefficients are obtained for atom-atom-atom and atom-atom-ion combinations. Here, we have taken only the singly ionized ions of group XII elements, \textit{viz.}, Zn$^+$, Cd$^+$, and Hg$^+$.

The evaluation of the dispersion coefficients for any combination is directly related to the dynamic dipole polarizabilities of atoms/ions, which we have obtained with the help of electric-dipole (E1) reduced matrix elements. The E1 reduced matrix elements are calculated by employing relativistic methods for the considered atoms and ions. Further, to ascertain the accuracy of the determined reduced matrix elements, we obtained using them the oscillator strengths for the leading transitions in atoms and ions which are then compared with the values reported by the National Institute of Standards and Technology (NIST) database \cite{NIST}. For the same purpose, the static dipole polarizabilities  in the ground states of atoms/ions are determined and compared with the measurements \cite{kompitsas1994,li2018,schwerdtfeger} wherever available. The outline of the present work is as follows: In Section~\ref{theory}, we give a brief overview of the theoretical methodology employed in the present work. The results are presented and discussed in Section~\ref{results}. The atomic units (a.u.) have been used throughout the manuscript. 

\section{THEORETICAL CONSIDERATION}~\label{theory}
%In this sections, the \textit{ab-initio} methods utilized to evaluate the two-dipole and three-dipole interaction terms has been discussed in details in respective sub-sections.
\vspace{-1.5cm}
\subsection{Dispersion coefficients}
The long-range interaction potential among the three atomic species in their ground states is expressed by following expression \cite{lee2013},
\begin{eqnarray}
V(\vec{R_{12}},\vec{R_{23}},\vec{R_{31}}) &=&  -\frac{C_6^{(12)}}{R^6_{12}}-\frac{C_8^{(12)}}{R^8_{12}}-\ldots-\frac{C_6^{(23)}}{R^6_{23}}-    \frac{C_8^{(23)}}{R^8_{23}}\nonumber\\
                                                            &-&\ldots-\frac{C_6^{(31)}}{R^6_{31}}-\frac{C_8^{(31)}}{R^8_{31}}-\ldots\nonumber\\
                                                            &+&(1+3\cos\theta_1\cos\theta_2\cos\theta_3)\frac{C^{123}_9}{R^3_{12}R^3_{23}R^3_{31}}.\label{pot}\nonumber\\
\end{eqnarray}
In this equation, the maximum contribution comes from the $C_n^{(ij)}$ parameters, which are the dispersion coefficients for two-dipole interactions. The three-dipole interaction terms  $C_n^{(ijk)}$, are the third-order correction to the ground-state energy and are known as the non-additive part of the potential. Therefore, it has different sign as compared to other interaction terms. Here, $i, j, k = 1, 2, 3$, are representing the three different atomic species. $R_{ij}$ is the inter-atomic distance between the two atomic systems. Further, the two $C_6^{(ij)}$ and $C_9^{(ijk)}$ coefficients can be expressed in terms of frequency-dependent dipole polarizability $\alpha(\iota\omega)$ as follows:
\begin{equation}
C_6^{(ij)}=\frac{3}{\pi}\int_0^{\infty}d\omega\alpha_{i}(\iota\omega)\alpha_{j}(\iota\omega), \label{ab-initioc6}
\end{equation}
and 
\begin{equation}
C_9^{(ijk)}=\frac{3}{\pi}\int_0^{\infty}d\omega\alpha_{i}(\iota\omega)\alpha_{j}(\iota\omega)\alpha_{k}(\iota\omega). \label{ab-initioc9}
\end{equation}
%to estimate these dispersion coefficients for group XII atoms and their ions, we employed \textit{ad-initio} methods.
 In order to calculate dispersion coefficients between the group XII atoms and their ions precisely, it is essential to use accurate values of frequency dependent dipole polarizabilities, whose evaluation method is discussed in the next section.  

\subsection{Dipole  Polarizability}

Following Ref.  \cite{arora2012}, the required dynamic dipole polarizability ($\alpha^{v}(\iota\omega)$) for an atomic system in its ground state $v$ can be written through three contributing terms:
\begin{equation}\label{dyn}
\alpha^{v}(\iota\omega)=\alpha^{v}_c(\iota\omega)+\alpha^{v}_{vc}(\iota\omega)+\alpha^{v}_{val}(\iota\omega),
\end{equation}
where, $\alpha^{v}_c(\iota\omega)$, $\alpha^{v}_{vc}(\iota\omega)$ and $\alpha^{v}_{val}(\iota\omega)$ denote the contributions from the core, valence-core and valence correlations, respectively. The core contributions are evaluated using the Dirac-Fock (DF) method for the closed-shell configurations \cite{safronova1998} by considering excitations from core to all valence orbitals. The valence-core $\alpha^{v}_{vc}(\iota\omega)$ contribution involves the transitions from core to occupied valence shells which are forbidden by the Pauli exclusion principle. Since for the considered  group XII elements the valence orbital is doubly filled, therefore the $\alpha^{v}_{vc}(\iota\omega)$ contribution for atoms is twice that of singly ionized ions. 

The dominant contribution to the total polarizability comes from the valence part, which can be calculated using the sum-over-states approach by the following expression:
\begin{eqnarray}\label{Eq.9}
& & \alpha^v_{val}(\iota\omega)=\frac{2}{3(2J_v+1)}
\nonumber \\
& & \times \sum_{p > N_c,p\neq v}^I \frac{(E_p-E_v)|\langle\psi_p||D||\psi_v\rangle|^2}{(E_p-E_v)^2+\omega^2}
                             \nonumber. \\
                             & &
\end{eqnarray}
In this equation, the summation is carried over all the possible intermediate states $p$ allowed by the electric dipole selection rules, and $N_c$ refers to the core orbitals. Static dipole polarizability can be calculated by substituting $\omega= 0$. 
This major contribution is again categorized in two parts, i.e., Main and Tail. The Main part includes the contributions primarily from the ground-state to low-lying allowed transitions, and in the case of the Tail part, it is from the ground state to the higher states of the atomic system. In Eq. \ref{Eq.9}, $\langle\psi_p||D||\psi_v\rangle|$ is the reduced dipole matrix element and is related to $\langle\psi_p|D|\psi_v\rangle$ through a  Clebsch–Gordan coefficient. For the Main part of the valence contribution, the required matrix elements are calculated using multi-configurational Dirac-Fock (MCDF) approximation and all-order (SD) method, respectively, for atoms and ions. The energies $E$s corresponding to the different states in these calculations are experimental and taken from the NIST database \cite{NIST}.
Further, the Tail part of the valence contributions is evaluated using the same expression, i.e., Eq. (5).  The matrix elements for the Tail part are calculated using the DF method for the ions and the required energy obtained through this method. However, this contribution for the atoms is neglected as these might not contribute significantly and for the sake of simplicity of the calculation.

\subsection{Evaluation of matrix element}~\label{matel}
 From the above discussion, we find that we need to evaluate the reduced dipole matrix elements for the determination of necessary dispersion coefficients among the group XII atoms and their ions. The desired wave functions for group XII atoms in their initial and final states are calculated through GRASP2K code \cite{jonsson2013} under the MCDF approach. In the MCDF approach, the atomic state wave function (ASFs) in the initial/final states can be expressed as a linear combination of various configurational state wave functions (CSFs), having the same total angular momentum and parity, i.e., 

\begin{equation}
|\Psi_v\rangle_{\rm MCDF}=\sum_{n=1}^Na_n|\Phi_n\rangle 
\end{equation}
here, $n$ denotes the number of CSFs, $a_n$ is the mixing coefficient of the CSF $|\Phi_n\rangle$. First, using Dirac-Coulomb Hamiltonian, the required expansion coefficients and single-particle orbital radial functions are determined by utilizing the multi-configuration self-consistent field calculations. Thereafter, by adding Breit and quantum electrodynamic corrections, the relativistic configuration interaction calculations are performed. To get accurate ASFs, we have used a large number of CSFs in the linear contribution, and in the final stage, we keep only those CSFs that have the corresponding mixing coefficient greater than 10$^{-3}$. 

Further, due to the suitability for the monovalent systems, we have utilized the singles and doubles excitation approximation in the all-order (SD) method \cite{Blundell,theory} to compute their wave functions. As compared to the MCDF approach, this method gives accurate results for them. However, for divalent systems, employing the SD method is rather challenging, and there are no codes available.

 The wave functions of the group XII ions in their respective state $v$, having the closed-core with a valence electron, is represented in the SD method by the following expression,
\begin{eqnarray}
|\Psi_v \rangle_{\rm SD} &=& \left[1+ \sum_{ma}\rho_{ma} a^\dag_m a_a+ \frac{1}{2} \sum_{mnab} \rho_{mnab}a_m^\dag a_n^\dag a_b a_a\right. \nonumber\\
  &&\left. + \sum_{m \ne v} \rho_{mv} a^\dag_m a_v + \sum_{mna}\rho_{mnva} a_m^\dag a_n^\dag a_a a_v\right] |\Phi_v\rangle,\nonumber \\
&&  
\label{sdmethod}
\end{eqnarray}
here, $|\Phi_v\rangle$ represents the Dirac Hartree Fock (DHF) wave function of the state $v$. $a_i$ and $a^\dag_i$ are the annihilation and creation operators. The indices $\{m,n\}$ and $\{a,b\}$ denote the virtual and core orbitals, respectively. $\rho_{ma}$ and $\rho_{mnab}$ are the single and double core coefficients. $\rho_{mv}$ and $\rho_{mnva}$ are the single and double valence excitation coefficients. The single-particle orbitals for the SD method have been constructed using the 70 B-spline functions, including a cavity of radius $R=220$ a.u.

Further for a particular transition of the concerned ions and atoms, with the help of calculated wave functions, the E1 reduced matrix elements are determined. 
%through the following expression:
\begin{comment}
\begin{eqnarray}\label{12}
	\langle\psi_p|D|\psi_v\rangle = \frac{\langle\phi_p|\tilde{D_{pv}}|\phi_v\rangle}{\sqrt{\langle\phi_p|\{1+\tilde{N_p}\}|\phi_p\rangle \langle\phi_v|\{1+\tilde{N_v}\}|\phi_v\rangle}}
	,\end{eqnarray}
here, $\tilde{D}_{pv} = \{1+S_p^\dagger\}e^{T^\dagger}Qe^T\{1+S_k\}$ and $\tilde{N}_{i=p,v} = \{1+S_i^\dagger\}e^{T^\dagger}e^T\{1+S_i\}$.\\ 
\end{comment}
To ascertain the accuracy of the matrix elements evaluated using the above-described methods, we also calculated the oscillator strengths and compared these with the values available from the NIST database~\cite{NIST}.  The following expression is used to evaluate the oscillator strengths \cite{pol-os}:  
\begin{equation}
f_{vp}=-\frac{303.756}{g_v\lambda_{vp}}\langle\psi_p||D||\psi_v\rangle|^2\label{em-os},
\end{equation}
here $f_{vp}$ is the  oscillator strength of the transition from state $v$  to excited intermediate states $p$. $g_v$ is the statistical weight and $\lambda_{vp}$ represents the transition wavelength in Angstrom.

\section{Results and Discussion}~\label{results}
In the present work, the theoretical calculations have been carried out to estimate the two-dipole ($C_6$)  and three-dipole ($C_9$) dispersion coefficients between the group XII atoms and their ions \textit{viz.}, Zn, Cd, Hg, Zn$^+$, Cd$^+$, and Hg$^+$. As explained earlier, to obtain these coefficients, the dynamic dipole polarizabilities and reduced matrix elements required are evaluated using the relativistic methods for both the considered atoms and ions. Further, using the obtained reduced matrix elements, the oscillator strengths corresponding to leading transitions and static dipole polarizabilities are determined. All the calculated results are discussed below in their respective sub-sections. 

\subsection{Oscillator strengths}
Using Eq.~\ref{em-os}, the oscillator strengths for the group XII atoms and ions are calculated. The results for few leading transitions of the considered atoms and ions are represented in Table~\ref{os}. 
As can be seen from the table that there is a close agreement between the present results and reported experimental values \cite{NIST} except for Hg$^+$. The exact reason for this discrepancy for Hg$^+$ is not known to us. While our values for Hg$^+$ matches completely with the theoretical results reported by Safronova and Johnson \cite{safronova2004}.  A thorough survey from the available literature and database (e.g., NIST) reveals that for the transitions of group XII atoms and their ions, only a limited number of oscillator strengths are reported. Also, we have not found any reported values of the oscillator strengths for any higher transitions of these atoms and ions. 
 However, we believe that our results, for which the previous values are not available for comparison, should also be reliable.  Thus, we can say that the reduced matrix elements of the considered atomic and ionic systems which are used to calculate the oscillator strengths, are also accurate as these are further utilized to evaluate the polarizabilities.  

\subsection{Static dipole polarizabilities}
With the help of determined reduced matrix elements, the static dipole polarizabilities of the group XII atoms and their ions are calculated. As mentioned above, that the sum-over-states approach has been utilized to obtain the polarizability. It should be mentioned here that for the ground state polarizability, only some initial transitions contribute the most; therefore, the truncation of the summation introduces only 1.5\% error in the calculation for ions and even less for the atoms. 

The calculated values of the static dipole polarizabilities for the considered atoms and ions are given in Table~\ref{pol}, where these are also compared with the previously reported theoretical \cite{gould,mitroy2010,chattopadhyay2015} and experimental \cite{kompitsas1994,li2018,schwerdtfeger} results.
%\textit{groupXII ions:} \\
 The comparison from the theoretical results of the polarizability for Zn$^+$ and Hg$^+$ \cite{mitroy2010} show very close agreement with the present values. On the other hand, the values reported by Chattopadhyay \textit{et al.} \cite{chattopadhyay2015} only for atoms and by Gould and Bucko~\cite{gould} for all ions as well as atoms show some difference in comparison with our results. This small difference arises due to the different methods used in the current, and the previous work \cite{gould,chattopadhyay2015}. Gould and Bucko\cite{gould} have utilized the time-dependent density functional theory with exchange kernels to obtain polarizabilities for atoms and ions. Whereas Chattopadhyay \textit{et al.} \cite{chattopadhyay2015} determined electric dipole polarizability of Zn, Cd, and Hg, by applying perturbed relativistic coupled-cluster (PRCC) theory. However, in the present study, the sum-over-sates approach has been employed, which allows us to use precise experimental values of energies wherever feasible \cite{NIST}. Thus, we believe that our determined values are quite reliable. Further, on comparing the present results with the available measurements \cite{kompitsas1994,li2018,schwerdtfeger}, we find that there are slight discrepancies with the relative difference in the range of 0.02-0.1 for ions, and in the case of atoms difference varies between 0.08 to 0.1.

\subsection{$C_6$ dispersion coefficient}
Further, with the help of calculated dynamic dipole polarizabilities, $C_6$ dispersion coefficients are evaluated for the combinations of two atoms as well as the atom and ion of group XII elements in their respective ground state. The calculation of $C_6$ for homonuclear and heteronuclear dimers of group XII atoms is especially useful for comparison purposes as there are some previous results available in the literature. This will ascertain the accuracy of our present method used to evaluate the dispersion coefficients for other combinations. The values of $C_6$ for homonuclear and heteronuclear dimers are listed in Table~\ref{c6-atom-atom1} and compared with other theoretical \cite{grycuk2012,qiao2012} and experimental results \cite{goebel199,goebel1996}. On comparison, we find that the values calculated by Qiao \textit{et al.} \cite{qiao2012} for all combinations are over-estimated by their method as compare to our calculations. On the other hand, Grycuk \textit{et al.} \cite{grycuk2012} have computed the dispersion coefficients only for homonuclear dimers, and their results match considerably with the present calculated values. One can also see from the Table~\ref{c6-atom-atom1} that the values of Qiao \textit{et al.} \cite{qiao2012} drastically differ from the results of Grycuk \textit{et al.} \cite{grycuk2012}. Further, we observe that our obtained coefficients for homonuclear combinations exhibit excellent agreement with the reported experiment results \cite{goebel199,goebel1996} for Cd-Cd and Hg-Hg dimers, where in the case of Zn-Zn, our result deviates from the measurement by $\sim $ 10\%. The overall good agreement of our results with the previously published works \cite{grycuk2012,qiao2012,goebel199,goebel1996} gives us confidence that the dispersion coefficients ($C_6$) obtained by us for different combinations between group XII atoms and their ions which are summarised in Table~\ref{c6-ion-atom} are quite reliable.

In addition, from the Tables ~\ref{c6-atom-atom1} and ~\ref{c6-ion-atom}, we notice that the relative magnitudes of the determined $C_6$ coefficients follow a specific trend, i.e., (Zn-X) $<$ (Hg-X) $<$ (Cd-X) (where X refers to the atom/ion of the group XII), that is probably due to the relativistic effects in the polarizability \cite {seth1997} which is different from the expected respective order of elements (Zn, Cd, Hg) in the group XII.

\subsection{$C_9$ dispersion coefficient}
In addition to $C_6$, the $C_9$ dispersion coefficients are calculated for the interaction of atom-atom-atom and atom-atom-ion combinations of the group XII elements, and these are listed in Tables~\ref{c9-atom-atom-atom} and~\ref{c9-atom-atom-ion}. Only for four combinations, we found the previously reported values for $C_9$, and these are given in Table~\ref{c9-atom-atom-atom}. However, for the rest of the other combinations, the $C_9$ coefficients which we are reporting here are the new ones. Now on comparison of our results with the available results, we find that values estimated by Qiao \textit{et al.} \cite{qiao2012} drastically differ with ours, similar to the $C_6$ case (as shown in Table~\ref{c6-atom-atom1}). In fact, Qiao \textit{et al.} \cite{qiao2012} have used Tang's \cite{tang1969} one-term approximation formula to evaluate their values for both the two- and three-dipole interaction terms, and thus our calculations seem more reliable than there. We have already seen from Table~\ref{c6-atom-atom1} that their values for $C_6$ were large as compare to our and the experimental results~\cite{goebel199,goebel1996}. Consequently, on the basis of good agreement between the present and previously reported other theoretical and experimental values \cite{grycuk2012,goebel199,goebel1996} (Table~\ref{c6-atom-atom1}) for the $C_6$ of homonuclear and heteronuclear dimers, gives us the confidence that the predicted three-dipole coefficients for various interactions are quite reliable. Here, for the $C_9$ coefficients,  we also observed the same trend as we have seen in the $C_6$ that the combinations which involve Cd atom in Tables~\ref{c9-atom-atom-atom} and~\ref{c9-atom-atom-ion}, have higher value of $C_9$ dispersion coefficients with respect to the other combinations. The reason for this could be the same as discussed in Section C. 

\section*{Conclusion}~\label{conclusion}
In the present work, we have investigated in a detailed manner the two-dipole ($C_6$) and three-dipole ($C_9$) interaction coefficients of the group XII elements (Zn, Cd, Hg) among themselves and with their singly ionized ions. For this purpose, we have calculated the reduced dipole matrix elements of atoms and ions by employing relativistic methods. Further, the oscillator strengths, dynamic and static dipole polarizability of these atoms, and ions are determined. Wherever possible, these results are compared with the other theoretical calculations, measurements, and the values from the NIST database, and overall good agreement is found. Thereafter, the calculated dynamic dipole polarizabilities are utilized to evaluate the $C_6$ and $C_9$ dispersion coefficients for various combinations. Most of these coefficients are reported for the first time in the literature. We hope that our values will be useful for the researchers for immediate use in various fields such as quantum information processing, designing better atomic clocks, and understanding the ion-atom hybrid mechanisms.    
\section*{Acknowledgements}

The authors, N.S. is thankful to the Ministry of Human
Resources and Development (MHRD), Govt. B. A. is thankfull to the SERB-TARE(TAR/2020/000189), New Delhi, India for research grant.  RS are thankful to the SERB-DST for the sanction of research grant No: CRG/2020/005597, New Delhi, Govt. of India.

\begin{table*}[h!]
\caption{Comparison of our calculated oscillator strengths ($f$) of the few leading transitions with the previously available values. The numbers in parentheses represent powers of 10.}
\label{os}
%\scalebox{0.7}{%
\begin{ruledtabular}
\begin{tabular}{lccccc}
Ion & Lower level & Upper level & Term & $f_{\rm present}$ & $f_{\rm previous}$ \\
\hline
Zn$^+$ & $4s$ & $4p$ &  $^2S_{1/2}$$\rightarrow$$^2P_{1/2}$ & 0.262(0)   &  0.246(0)$^a$ \\
Zn$^+$ & $4s$ & $4p$ &  $^2S_{1/2}$$\rightarrow$$^2P_{3/2}$ & 0.535(0)  & 0.501(0)$^a$  \\
Zn$^+$ & $4s$ & $5p$ &  $^2S_{1/2}$$\rightarrow$$^2P_{1/2}$ & 0.001(0)   &  - \\
Zn$^+$ & $4s$ & $5p$ &  $^2S_{1/2}$$\rightarrow$$^2P_{3/2}$ & 0.001(0)   &  - \\
Cd$^+$ & $5s$ & $5p$ &  $^2S_{1/2}$$\rightarrow$$^2P_{1/2}$ & 0.252(0)   &  0.244(0)$^a$ \\
Cd$^+$ & $5s$ & $5p$ &  $^2S_{1/2}$$\rightarrow$$^2P_{3/2}$ & 0.536(0)   &  0.521(0)$^a$ \\
Cd$^+$ & $5s$ & $6p$ &  $^2S_{1/2}$$\rightarrow$$^2P_{1/2}$ & 0.001(0)   &  -\\
Cd$^+$ & $5s$ & $6p$ &  $^2S_{1/2}$$\rightarrow$$^2P_{3/2}$ & 0.0005(0)   & -\\
Hg$^+$ & $6s$ & $6p$ &  $^2S_{1/2}$$\rightarrow$$^2P_{1/2}$ & 0.214(0)   & 0.42(0)$^a$\\
Hg$^+$ & $6s$ & $6p$ &  $^2S_{1/2}$$\rightarrow$$^2P_{3/2}$ & 0.507(0)   & 0.98(0)$^a$\\
Hg$^+$ & $6s$ & $7p$ &  $^2S_{1/2}$$\rightarrow$$^2P_{1/2}$ & 0.047(0)   & 0.083(0)$^a$\\
Hg$^+$ & $6s$ & $7p$ &  $^2S_{1/2}$$\rightarrow$$^2P_{3/2}$ & 0.023(0)   & 0.014(0)$^a$\\
\hline
Atom & Upper level & Lower level & Term & $f_{\rm present}$ &   $f_{\rm previous}$ \\
\hline
Zn      & $4s^2$  & $4s4p$    & $^1S_0$$\rightarrow$$^3P_1$ & 0.164(-4) & - \\
Zn      & $4s^2$  & $4s4p$    & $^1S_0$$\rightarrow$$^1P_1$ & 1.46(0) & 1.46(0)$^a$ \\
Zn      & $4s^2$  & $4s5p$    & $^1S_0$$\rightarrow$$^3P_1$ & 0.654(-5) & - \\
Zn      & $4s^2$  & $4s5p$    & $^1S_0$$\rightarrow$$^1P_1$ & 0.089(0) & 0.089(0)$^a$ \\
Cd      & $5s^2$  & $5s5p$    & $^1S_0$$\rightarrow$$^3P_1$ & 0.543(-3) & 0.194(-2)$^a$ \\
Cd      & $5s^2$  & $5s5p$    & $^1S_0$$\rightarrow$$^1P_1$ & 1.26(0) & 1.2(0)$^a$ \\
Cd      & $5s^2$  & $5s6p$    & $^1S_0$$\rightarrow$$^3P_1$ & 0.289(-3) & - \\
Cd      & $5s^2$  & $5s6p$    & $^1S_0$$\rightarrow$$^1P_1$ & 0.887(0) & - \\
Hg     & $6s^2$  & $6s6p$    & $^1S_0$$\rightarrow$$^3P_1$ & 0.141(-1) & 0.243(-1)$^a$ \\
Hg     & $6s^2$  & $6s6p$    & $^1S_0$$\rightarrow$$^1P_1$ & 0.119(0) & 0.115(0)$^a$ \\
Hg     & $6s^2$  & $6s7p$    & $^1S_0$$\rightarrow$$^3P_1$ & 0.197(-3) & - \\
Hg     & $6s^2$  & $6s7p$    & $^1S_0$$\rightarrow$$^1P_1$ & 0.012 & - \\
\end{tabular}%}
\end{ruledtabular}
$^a$NIST\cite{NIST}
\end{table*}

\begin{table*}[h!]
\caption{Comparison of our calculated static dipole polarizabilities $(\alpha^v(0))$ (in a.u.) for the considered group XII atoms and their singly ionized ions in the ground states with the available measurements and other theoretical calculations. }
\label{pol}
%\scalebox{0.7}{%
\begin{ruledtabular}
\begin{tabular}{lcccc}

Ion &  State & Present Calculations & Other Calculations  & Measurements   \\
\hline
Zn$^+$       &    4s    & 18.05&17.9$^a$,18.8$^b$& 15.4$^c$ \\
Cd$^+$       &    5s  & 23.41& 23.1$^a$&25.2(6)$^d$   \\
Hg$^+$       &    6s    & 19.07&17.5$^a$,19.36$^b$& - \\
\hline
Atom &  State & Present Calculations & Other Calculations & Measurements   \\
\hline
Zn     &    $4s^2$    &35.33 & 38.4$^a$,38.71$^e$& 38.67$^f$\\
Cd     &    $5s^2$    &49.61 & 46.7$^a$,48.2$^e$& 48.2$^f$\\
Hg      &    $6s^2$     &28.65 &33.5$^a$,33.63$^e$& 33.91$^f$ \\
\end{tabular}%}
\end{ruledtabular}
$^a$Ref.\cite{gould}, $^b$Ref.\cite{mitroy2010}, $^c$Ref.\cite{kompitsas1994}, $^d$Ref.\cite{li2018}, $^e$Ref.\cite{chattopadhyay2015},$^f$Ref.\cite{schwerdtfeger}.
\end{table*}

\begin{table*}[h!]
\caption{ The calculated values of the two-dipole interactions ($C_{6}$) between the two atoms (heteronuclear and homonuclear) of the group XII in their respective ground state and comparison with the previously reported results.
\label{c6-atom-atom1}}
\begin{ruledtabular}
\begin{tabular}{lccc}
           Dimers  & Present Calculation & Other Calculations & Measurements  \\ 
\hline
Zn-Zn & 225 & 298$^a$,359$^b$ & 257.5$^c$ \\
Zn-Cd  & 332 & 494.5$^b$ & - \\
Zn-Hg  & 229 & 368.7$^b$ & - \\
Cd-Cd & 493 & 487$^a$,686$^b$ & 466$^d$ \\
Cd-Hg & 347 & 515.8$^b$ & - \\
Hg-Hg  & 258 & 272$^a$,392$^b$ & 255$^c$ \\
\end{tabular}%}
\end{ruledtabular}
$^a$Ref. \cite{grycuk2012}, $^b$Ref. \cite{qiao2012}, $^c$Ref. \cite{goebel1996}, $^d$Ref. \cite{goebel199}.
\end{table*}

\begin{table*}[h!]
\caption{The calculated values of the two-dipole interactions ($C_{6}$) between the atoms and singly ionized ions of the group XII in their respective ground state. \label{c6-ion-atom}}
\begin{ruledtabular}
\begin{tabular}{lccc}
 X& Zn-X & Cd-X   & Hg-X     \\
\hline
Zn$^+$ &124&185& 132\\
Cd$^+$ &172 & 259& 190\\
Hg$^+$	&171 & 262 & 201\\
\end{tabular}%}
\end{ruledtabular}
%\vspace{-1.5cm}
%\end{table*}

%\begin{table*}[h!]
 \caption{ The calculated values of the three-dipole interactions ($C_{9}$) among the three atoms of the group XII in their respective ground states and comparison with the previously reported results. \label{c9-atom-atom-atom} }
\begin{ruledtabular}
\begin{tabular}{lccc}
              X     &    Zn-Zn-X & Cd-Cd-X   & Hg-Hg-X          \\
\hline

Zn & 5667& 11650 & 4776 	\\
&8940$^a$&16239$^a$&- \\
Cd &8118 &16751& 6971 \\
&12023$^a$&22033$^a$&- \\                       
Hg &5124 & 10698   & 4666 	\\
&-&-&- \\                   \end{tabular}%}
\end{ruledtabular}
$^a$Qiao \textit{et al.}\cite{qiao2012}
%\vspace{-3cm}
%\end{table*}

%\begin{table*}[h!]
\caption{ The calculated values of the three-dipole interactions ($C_{9}$) among the two atoms and singly ionized ions of the group XII atoms in their respective ground states.\label{c9-atom-atom-ion} }
\begin{ruledtabular}
\begin{tabular}{lccc}
                 X   &Zn-Zn-X &  Cd-Cd-X   & Hg-Hg-X        \\
\hline
Zn$^+$&2994& 6185 &  2588  \\
Cd$^+$ &3960 & 8227 & 3524  \\
Hg$^+$	&3592 & 7559 & 3401  \\

\end{tabular}%}
\end{ruledtabular}
\end{table*}

%\bibliography{ref.bib}

\end{document}